\documentclass{ametsocV6.1}



\usepackage{amsmath,amsfonts,amssymb,bm}
\usepackage{mathptmx}

\usepackage[T1]{fontenc}
\usepackage{color,soul}

\usepackage{textcomp}
\usepackage{nicefrac} 
\usepackage{xfrac}
\usepackage{mathtools}
\usepackage{newtxmath}
\usepackage{newtxtext}

\title{Physically Explainable Deep Learning for Convective Initiation Nowcasting Using GOES-16 Satellite Observations}
\begin{document}
\authors{Da Fan$^a$\correspondingauthor{Da Fan, dxf424@psu.edu}, Steven J. Greybush$^a$, David John Gagne II$^b$, and Eugene E. Clothiaux$^a$}\
\affiliation{\aff{a}Department of Meteorology and Atmospheric Science, The Pennsylvania State University, University Park, Pennsylvania\\
\aff{b}National Center for Atmospheric Research, Boulder, Colorado}
%
%
\abstract
{Convection initiation (CI) nowcasting remains a challenging problem for both numerical weather prediction models and existing nowcasting algorithms. In this study, object-based probabilistic deep learning models are developed to predict CI based on multichannel infrared GOES-R satellite observations. The data come from patches surrounding potential CI events identified in Multi-Radar Multi-Sensor Doppler weather radar products over the Great Plains region from June and July 2020 and June 2021. An objective radar-based approach is used to identify these events. The deep learning models significantly outperform the classical logistic model at lead times up to 1 hour, especially on the false alarm ratio. Through case studies, the deep learning model exhibits the dependence on the characteristics of clouds and moisture at multiple levels. Model explanation further reveals the model's decision-making process with different baselines. The explanation results highlight the importance of moisture and cloud features at different levels depending on the choice of baseline. Our study demonstrates the advantage of using different baselines in further understanding model behavior and gaining scientific insights.
}

\maketitle

\section{Introduction}

Convective initiation (CI) remains a significant and challenging forecasting problem within the meteorological community. Accurately predicting the location and onset times of convection remains difficult for both empirical and numerical weather prediction (NWP) models \citep[e.g.,][]{Mecikalski2015, Lawson2018, Cintineo2020}. The failure to forecast CI causes delayed warnings of convective hazards like heavy rainfall, hail, and tornadoes, and disruptions to outdoor activities and travel \citep[][]{Brooks2003, Brooks2008, Dixon2011}. Given its socioeconomic impacts, more accurate and timely CI nowcasts and a more thorough understanding of physical processes underlying CI are needed.

Multiple satellite-based algorithms have been developed to make use of cloud characteristics to enhance the forecast skill of CI \citep{Roberts2003,Mecikalski2006,Mecikalski2010,Sieglaff2011,Walker2012}. The University of Wisconsin Convective Initiation (UWCI) nowcasting algorithm \citep{Sieglaff2011} was developed to nowcast CI based on box-average cloud-top characteristics evident within Geostationary Operational Environmental Satellite (GOES) observations in areas not covered by anvil clouds. The Satellite Convection Analysis and Tracking, version 2, (SATCASTv2) algorithm was developed by \cite{Walker2012} to track cumulus clouds and nowcast the probability of CI in the cloud objects. Cloud-top features, like cloud-top cooling rate and hydrometeor phase, were employed to predict CI \citep{Sieglaff2011,Walker2012}. However, the performances of these algorithms were hindered by high false alarm ratios \citep{Sieglaff2011,Walker2012}, even with the aid of environmental conditions from NWP model simulations \citep{Mecikalski2015, Apke2015}. Another issue is that a substantial number of CI events blocked by thick cirrus clouds were ignored in these algorithms focusing on growing cumulus clouds \citep{Walker2012}. Spatial variations of cloud-top features surrounding potential CI events were not used to improve the forecast skill, probably due to the coarse resolution of the past satellite observations.

Fine-resolution infrared observations are now available every few minutes from the current generation of geostationary satellites. These high spatiotemporal resolution satellite observations have increased our ability to better represent cloud-top characteristics associated with convection \citep[e.g.,][]{Senf2017, Apke2018,Fan2022}. Enhanced spatial patterns of environmental information have been found critical for predicting severe hailstorms \citep{Gagne2019} and tornadoes \citep{Lagerquist2019} using deep learning methods. However, spatial features in these new satellite observations have not yet received much attention for increasing CI forecast skill.

Machine learning (ML) and deep learning methods have recently gained popularity as a powerful tool in a variety of weather and climate applications, such as convective weather forecasting \citep{McGovern2019, Gagne2019}, tropical-cyclone intensity prediction \citep{Wimmers2019}, and subseasonal forecasting of tropical-extratropical circulation teleconnections \citep{Mayer2021}. With an ability to encode complex spatial features \citep{Lecun2015}, deep learning improves forecast skill through encoding relationships inherent within immense quantities of data. \cite{Gagne2019} showed that a convolutional neural network (CNN), a popular deep learning method, performed well on predicting severe hailstorms based on simulated thermodynamic fields. \cite{Lee2020} developed a CNN for detecting convective regions from satellite observations with improved accuracy. \cite{Lagerquist2021} further demonstrated that deep learning provides skillful forecasts of the spatial coverage of convection at lead times up to 120 minutes using infrared satellite data. The successful applications of deep learning to detect and forecast convection with satellite observations hold promise for further using deep learning to improve convective initiation forecasting.

Despite the increase of ML’s successful applications in meteorology, it is often criticized by forecasters and domain scientists as a “black box” technique because of our inability to readily interpret its decision-making process in physical terms. Thus, explainable artificial intelligence (XAI) has received a lot of attention in both the meteorology and ML communities \citep{Olah2017, Lipton2018, McGovern2019, Toms2020, Molnar2020}. XAI encapsulates and approximates intricate relations between inputs and model predictions inherent in the decision-making process, enabling domain scientists to gain trust in the model, as well as understand its limitations. This facilitates application of the model to ideal scenarios \citep{McGovern2019} and withholding it from inappropriate ones. XAI is becoming increasingly important as ML methods outperform current NWP models in some applications.

Successful applications of XAI include \cite{Toms2020}, who identified the spatial patterns for two dominant modes of El Nino variability using layer-wise relevance propagation (LRP) and backwards optimization, two model explanation methods. \cite{Mayer2021} demonstrated that neural networks are able to identify tropical hot spots that are important for subseasonal predictions in the North Atlantic through model explanations using LRP. \cite{Mamalakis2022} objectively assessed the performance of different explanation methods on a large benchmark dataset and discussed their reliability and limitations compared to the ground truth.

The purpose of this study is to characterize nowcasting skill of CI obtained through two ML models trained on GOES-16 satellite infrared observations and to explore the radiative features that lead to skill in forecasting CI through model explanation and visualization. A CNN model is optimized and evaluated against logistic regression, a classical statistical method. False positive prediction is a crucial issue in previous CI forecasting algorithms \citep{Mecikalski2015, Apke2015}, so particular attention is paid to this challenge. The rest of this paper is organized as follows. Section 2 describes CI identification and data preprocessing. Section 3 describes model architectures, optimization, evaluation, and explanation methods. Section 4 evaluates the CNN and logistic regression models through performance statistics and case examples. Section 5 explains radiative features in the decision-making process of the CNN with different choices of the baseline. Section 6 presents the main findings and limitations of the study, and includes concluding remarks.

\section{Data}

We use GOES-16 Advanced Baseline Imager (ABI) data to generate predictors for CI events obtained from the Multi-Radar Multi-Sensor \citep[MRMS;][]{Lakshmanan2006, Lakshmanan2007} dataset. This study focuses on the Great Plains in the United States (Fig. 1), following \cite{Apke2015} and \cite{Walker2012}, because of the importance of CI to this region and the availability of dense radar observations within it.

\subsection{CI definition and identification}
\begin{figure*}[t]
\centerline{\includegraphics[width=25pc,angle=0]{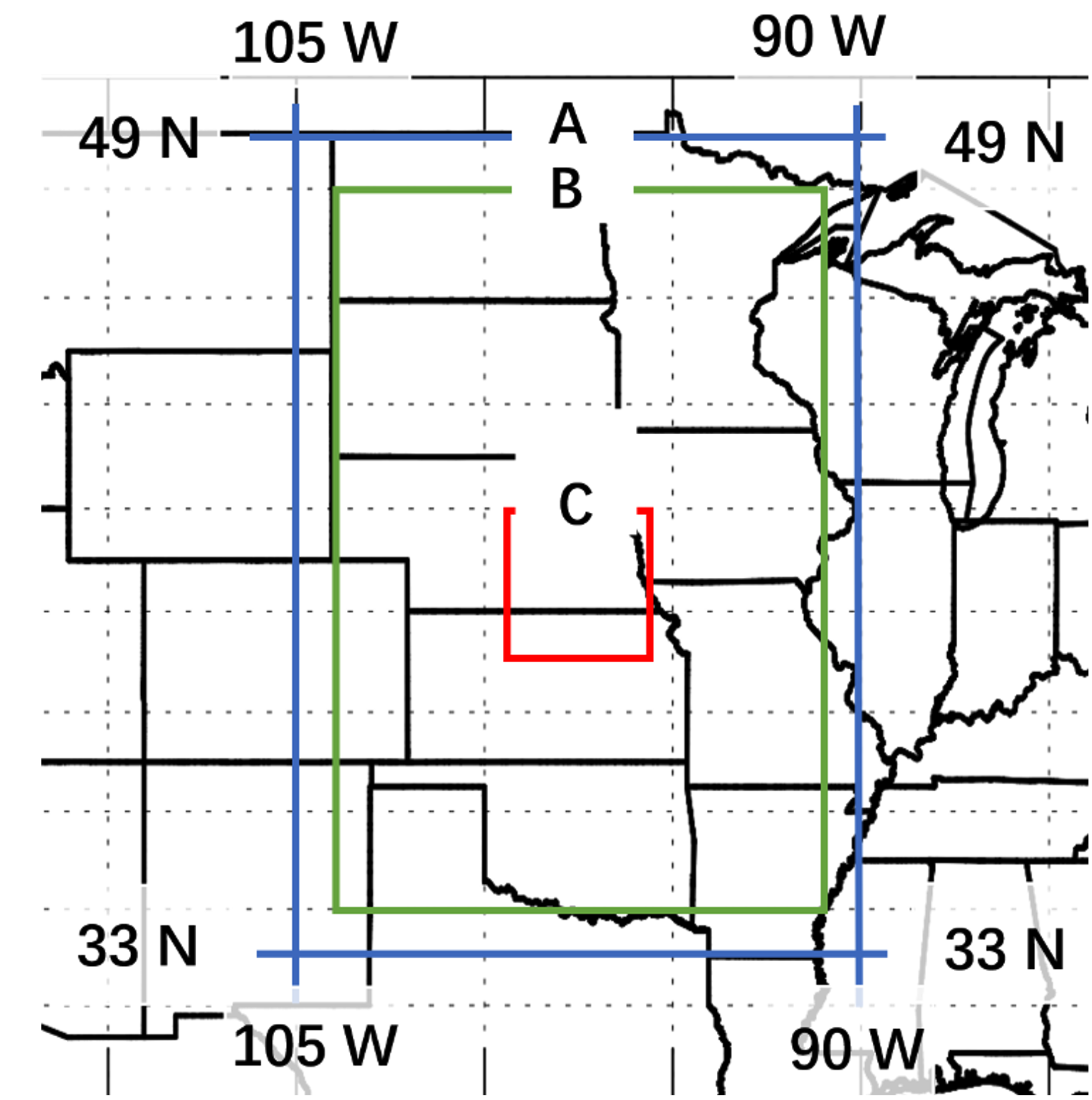}}
\caption{Total area of study (blue box labeled with A), track-corrected area of study (green box labeled with B, and ranging from 91°W to 104°W and from 34°N to 48°N), and validation domain for manually identified CI true clusters (red box labeled with C).}\label{fig1}
\end{figure*}

Radars used to produce the MRMS dataset make azimuth scans at a number of elevation angles \citep{Lakshmanan2006}. These radar data are then interpolated onto a uniform grid for postprocessing. Composite reflectivity, defined as the maximum reflectivity in a column on a 0.01° spatial grid and 2-min temporal grid in the MRMS dataset, is used for storm tracking and CI identification. Using a column-max value mitigates the impacts of terrain blocking radar beams and increases the probability of detecting convective cells in their early stages as hydrometeors form at higher levels \citep[e.g.,][]{Matthee2014, Apke2015, Senf2017, Henderson2021}. A radar reflectivity threshold of 35 dBZ is employed to distinguish between convective and nonconvective regions \citep{Mecikalski2006, Kain2013}. Pixels are defined as being part of a CI cluster when the following conditions \citep[adapted from][]{Colbert2019} are met:

(1) composite reflectivity $\geq$ 35 dBZ;

(2) in the preceding 11 min, no points within 15 km exhibit composite reflectivity $\geq$ 35 dBZ; and

(3) in the preceding 30 min, no points within 5 km exhibit composite reflectivity $\geq$ 35 dBZ.

The first condition identifies pixels associated with convection either just initiated or advected to the pixel from the surroundings. The second and third conditions eliminate pixels related to preexisting convection. Storm tracking and CI cluster identification are done in three steps. First, storm cells are identified and clustered into tracks iteratively in time using the w2segmotionll algorithm \citep{Lakshmanan2010}, which is part of the Warning Decision Support System–Integrated Information (WDSS-II) suite of algorithms. Combined K-means and enhanced watershed methods \citep{Lakshmanan2009} are used for this purpose. Second, the tracks are corrected by a modified best-track algorithm \Citep{Lakshmanan2015} using post-event tracking, which fits the storm cells to a best-fit Theil-Sen trajectory and removes falsely truncated tracks. Third, the first storm cells within the final tracks are identified as CI clusters. CI clusters are validated through two examples selected from the area C in Fig. 1 (See the video in the online supplemental material).

\subsection{Feature engineering}
An object-based forecasting method is designed to identify localized environments within which to predict CI, thereby largely reducing the data volume. Both CI and non-CI events must be identified for the dataset. CI events are 48-km by 48-km square patches centered on at least one CI cluster, whereas most ($\sim$91\%) non-CI events, called Near-Miss (NM) events, are 48-km by 48-km square patches that are nearest neighbors to CI events and contain no CI cluster of their own. The rest ($\sim$9\%) of the non-CI events, called RandoM (RM) events, are 48-km by 48-km square patches randomly extracted across the Great Plains area and neither contain a CI cluster nor are a nearest neighbor to a patch that does. To avoid the impacts of class imbalance on model performance \citep{Ukkonen2019}, non-CI events, the majority class, are undersampled to produce a balanced dataset that consists of 58\% CI, 38\% NM, and 4\% RM events. We use binary labels to classify these events, so that 1 indicates a CI event and 0 a non-CI event. Our entire dataset consists of 94,618 samples. 45,077, and 19,320 samples from June and July 2020 are for training and validation, respectively, whereas 30,221 samples collected in June 2021 are for testing.
\begin{table}[t]
\caption{Central wavelength, mean, and standard deviation over the training dataset of infrared brightness temperatures from GOES-16 ABI used as predictors.}\label{table1}
\centering
\begin{tabular*}{0.7\hsize}{@{\extracolsep\fill}cccc@{}}
\topline
Channel& Central wavelength ($\mu m$)& Mean (K) & Standard deviation (K)\\
\midline
\ CH8 & 6.2 & 226.7 & 11.3\\
\ CH9 & 6.9 & 231.2	& 14.9\\ 
\ CH10 & 7.3 & 235.3 & 18.0\\
\ CH11 & 8.4 & 247.3 & 29.2\\
\ CH12 & 9.6 & 239.8 & 14.6\\
\ CH13 & 10.3 & 248.8 & 30.2\\
\ CH14 & 11.2 & 247.6 & 30.1\\
\botline
\end{tabular*}
\end{table} 
Predictors for CI and non-CI events are channel 8-14 infrared brightness temperatures (BTs; Table 1) from GOES-16 ABI with $\sim$2-km native horizontal resolution available every 5 minutes over the continental United States (CONUS). Predictors are extracted from the 48-km by 48-km square patches (i.e., around 24 columns by 24 rows in a GOES-16 ABI image) at lead times from 60 minutes down to 10 minutes before the occurrence time of an event. As GOES-16 ABI views clouds across CONUS slantwise, their surface referenced latitudes and longitudes in the database differ from their vertically projected latitudes and longitudes, with the difference largest for the highest altitude clouds. This displacement, called parallax error, is comparable to the scale of the clouds during thunderstorm initiation \citep{Zhang2019} and thus not negligible. Following \cite{Zhang2019}, parallax errors are corrected using the cloud-top height (ACHA) product of GOES-16 to improve the quality of cloud locations.

The depth and performance of CNNs are largely limited by the size of the input \citep{Thambawita2021, Sabottke2020}, with small-size inputs usually leading to shallow CNNs with limited ability to encode complex spatial features. Thus, GOES-16 ABI BTs are remapped to a 1.5-km mesh, so that the 48-km by 48-km square patches contain 32×32 input BT values. Each predictor set of channel BTs is standardized using its mean and standard deviation (Table 1) to a set of values with zero mean and standard deviation of one prior to being fit by the two ML models.

\section{Methods}
This section introduces the architecture of the logistic regression and CNN models, a hyperparameter optimization method, and model explanation approaches. 
\subsection{Logistic regression}

Logistic regression is a nonlinear transformation with a sigmoid function applied to the weighted sum of the predictors ($x_i$):
\begin{equation}
p=\frac{1}{1+e^{-z}}\text{, where } z=\beta_{0}+\sum_{i=1}^{N}\beta_{i}x_{i}\text{,}
\end{equation}
$p$ is the prediction in the range between zero and one, $N$ is the total number of predictors, $\beta_{i}$ is the $i^{th}$ learned weight, and $\beta_{0}$ is the bias term. Predictions from logistic regressions are often used to estimate probabilities for classification problems with monotonic relationships between predictors and predictands. The model weights are iteratively adjusted by minimizing the binary cross-entropy
\begin{equation}
C\sum_{j=1}^{M}[y_{j}\log_2(p_{j}) + (1 - y_{j})\log_2(1 - p_{j})] + \lambda \sum_{j=1}^{M} | \beta_{j}| + \frac{1-\lambda}{2} \sum_{j=1}^{M} |\beta_{j}|^2 \label{eq:2}
\end{equation}
between the true labels ($y_{j}$) and the predictions ($p_{j}$), where the two additional terms, known as elastic-net penalties, are for regularization, $M$ is the number of samples, $C$ is the inverse of the regularization strength, and $\lambda$ is the mixing parameter that controls the strengths of the two regularization terms. The second term in Eq.(\ref{eq:2}) is known as the lasso penalty, or $L_{1}$ regularization, and rewards small weights by penalizing the sum of absolute values of the weights \citep{Tibshirani1996}. The third term is known as the ridge penalty, or $L_{2}$ regularization, and reduces the impacts of multicollinearity, i.e., correlations between predictors, by adding additional penalties to large weights \citep{Hoerl1988}. 

Despite being a simple ML model, logistic regression performs well on some problems in weather forecasting, like distinguishing between lightning and non-lightning days \citep{Bates2018} and predicting CI using satellite observations \citep{Mecikalski2015}. In our study, the baseline logistic model feeds flattened GOES-16 predictors into a logistic regression to predict the probability of CI. The logistic model is implemented in version 0.20 of the scikit-learn \citep{PedregosaFABIANPEDREGOSA2011} library.

\subsection{Residual neural network}

\begin{figure*}[t]
\centerline{\includegraphics[width=\textwidth,angle=0]{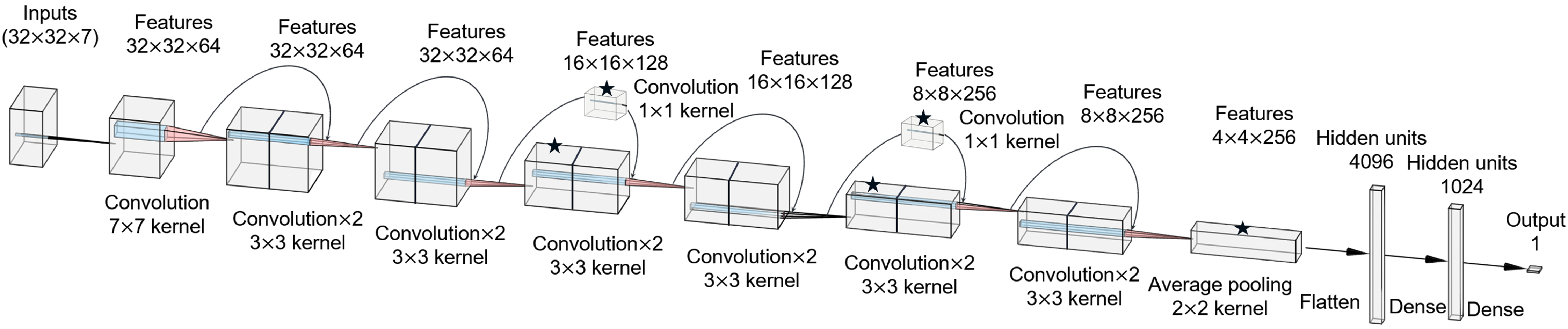}}
\caption{Schematic of the ResNet architecture. The curves connecting the input and output of residual blocks indicate residual layers. The blocks marked by a star involve downsampling of features from the preceding layer. The architecture parameters and components shown produced the highest validation score during the hyperparameter optimization.}\label{fig2}
\end{figure*}

We also trained a deep CNN architecture to predict the probability of CI. Deep neural networks often perform better than shallow networks by encoding spatial features across multiple scales through higher-order abstraction. A drawback of deep learning networks is potential difficulty with convergence resulting from vanishing gradients during the optimization process \citep{Glorot2010}. Gradients decrease exponentially as they are propagated back to the early layers so that optimization of the weights and biases in the early layers becomes problematic during training. To avoid this issue, we used the residual neural network (ResNet) architecture \citep{He2016} with its residual connections. In this architecture, the output of earlier layers is added to the output of later layers to preserve high-resolution information and thus preserve the gradients. The ResNet method performed well in medium-range weather forecasting in the WeatherBench challenge, and its skill is comparable to the baseline physical model at similar resolution \citep{Rasp2021}. 

A schematic of the ResNet architecture used in this study is shown in Fig. 2. For the first layer, a single two-dimensional (2D) convolutional block extracts features from the inputs using a kernel size of (7, 7) to broaden the view of the field. The convolutional block is defined as a sequence of 2D convolution layer → Leaky rectified linear (ReLU; max[0,x]) activation → Batch normalization → Dropout. The 2D convolution layer extracts spatial features from the inputs by transforming them through a number of filters, which are small patches of weights and biases. Then, the Leaky ReLU activation function is applied to the feature maps to preserve only positive signals \citep{Lecun2015}. Batch normalization \citep{Ioffe2015} is subsequently used to rescale the values of the features in order to maintain a more stable structure during training and to enable faster convergence of model errors with higher learning rates. Dropout regularization then randomly sets the values of certain features to zero with a fixed probability to prevent overfitting \citep{Srivastava2014}.

After the first convolutional block, there are six residual blocks with each block consisting of two 2D convolutional blocks with a 3×3 kernel and a residual layer. The residual layer adds the features of the preceding layer to the features of the residual block to increase the magnitude of the gradients. In the third and fifth residual blocks, the features from the preceding layer are downsampled via a residual layer with a 1×1 kernel and a stride of 2 and a convolutional block with a 3×3 kernel and a stride of 2. The number of convolutional filters increases by a factor of two from the second to third and fourth to fifth residual blocks. This is done to offset the loss of information from a decrease in spatial resolution at these steps. The spatial resolutions of the feature maps decrease with increasing depth, and the feature maps evolve to contain different levels of abstraction. The average pooling layer reduces the dimensions of the feature maps by a factor of 2 via a convolution with a 2×2 kernel and a stride of 2, thereby refining the features used for prediction. The resulting features are then flattened into a one-dimensional feature vector. The vector is then condensed through two fully connected dense layers. Each feature of a dense layer is a weighted sum of features from the previous layer. The outputs of the final dense layer are transformed through a sigmoid activation function into the probability of CI.

We trained the ResNet model using binary cross-entropy as the loss function, and we used area-under-curve (AUC) scoring as the metric to track model performance on the validation data during training. An Adam optimizer was used with an initial learning rate. The learning rate was decreased by a factor of two after validation losses did not decrease across three training epochs. We terminated training when the validation losses did not decrease across ten training epochs. We built the ResNet model using the Keras library \citep{chollet2015keras} with a Tensorflow low-level backend \citep{Abadi2016}.
\subsection{Model optimization and evaluation} 

To find the optimal configuration for both the logistic regression model and the ResNet, we performed a guided search over a range of hyperparameters using Earth Computing Hyperparameter Optimization (ECHO: \url{https://doi.org/10.5281/zenodo.7787022}). The search is based on the Tree-structured Parzen Estimator (TPE) sampler, which samples the next hyperparameters based on the ranking information of previous experiments. We performed 200 hyperparameter searches and selected the model with the highest AUC score on the validation data. The selected hyperparameters, search space, and optimal values are shown in Table S2.

We then evaluated the optimized logistic regression and ResNet models on the testing dataset. Most performance metrics were derived from the relationship between CI/non-CI observations and binary deterministic predictions (“yes”/”no”) converted from probabilistic forecasts using a probability threshold. The four possible outcomes are: 1) hits: correctly forecast CI occurrences, 2) false alarms: CI forecast where no CI occurred, 3) correct negatives: correctly forecast non-CI occurrences, 4) misses: non-CI forecast where CI occurred. Commonly used metrics for deterministic forecasts include probability of detection [POD; h/(h+m)], probability of false detection [POFD; f/(f+c)], false alarm ratio [FAR; f/(f+h)], success ratio [SR; h/(h+f)], frequency bias [(h+f)/(h+m)] and critical success index [CSI; h/(h+m+f)], where h, f, c, and m are the frequency of hits, false alarms, correct negatives, and misses, respectively. The probability thresholds were selected to maximize the CSI for both the logistic regression and ResNet models. The models were individually trained using the same architecture at lead times from 60 minutes to 10 minutes prior to the event, in steps of 10 minutes. Then, the models were evaluated and compared to each other.

The final skill score in this study is the Brier skill score \citep[BSS;][]{Wilks2019}. The BSS is a measure of the improvement of the forecast skill relative to climatology based on how well the probabilistic forecast agrees with the observed event frequency. The BSS is decomposed into two terms with a scaling factor:
\begin{equation}
 BSS = \frac{Resolution-Reliability}{Uncertainty} = \frac{\frac{1}{N}\sum_{k=1}^{K} n_{k}(y_{k}-\overline{y})^2-\frac{1}{N}\sum_{k=1}^{K} n_{k}(p_{k}-y_{k})^2}{\overline{y}(1-\overline{y})}
\end{equation}
where $N$ is the total number of samples, $K$ is the number of bins, $n_{k}$ is the number of samples in the $k^{th}$ probability bin, $y_{k}$ is the conditional event frequency given the probability in the $k^{th}$ bin, $p_{k}$ is the mean probability in the $k^{th}$ bin, and $\overline{y}$ is the climatological event frequency. The two terms in the numerator are known as the resolution and the reliability, while the scaling factor in the denominator is known as the uncertainty. The resolution measures the difference between conditional event frequencies and the observed climatological frequency, whereas the reliability measures how close the forecast probabilities are to the observed frequency at the corresponding probabilities. The uncertainty term rescales the BSS score based on the class proportion. A BSS score of 1 represents a perfect model, whereas a score less than 0 indicates the model is worse than climatology.

\subsection{Model explanation}
To explain the encoding underlying the complex structures of the ResNet model, we employed a model-agnostic method called SHapley Additive exPlanations \citep[SHAP;][]{Lundberg2017}. SHAP explains individual predictions as a game played by features and fairly distributes the payout among the features \citep{Molnar2020}. A player of the game is an individual feature value or a group of feature values. To explain an image input, pixels are grouped into superpixels and contributions to the prediction are distributed among them. The SHAP method estimates the Shapley value of each feature as its contribution to the prediction. The Shapley value is the only explanation method with a solid theory \citep{Young1985} that satisfies symmetry, local accuracy (also known as additivity), and consistency, properties not pertinent to other XAI methods. 

\section{Performance evaluation}
We break down model performance into two parts. First, we present summary statistics of model performance to provide an overview of how each model performed. Then, we present examples of what the models got right and wrong.

\subsection{General performance}
\begin{table}[t]
\caption{Performance scores of the baseline logistic regression and ResNet models on the testing dataset at a 10-min lead time. The probability threshold was selected to generate the optimal CSI for the validation dataset.}\label{table2}
\centering
\begin{tabular*}{0.7\hsize}{@{\extracolsep\fill}ccccc@{}}
\topline
Model & AUC & CSI & FAR & BSS\\
\midline
\ Logistic Regression & 0.758 & 0.666 & 0.302 & 0.238\\
\ ResNet & 0.855 & 0.723 & 0.236 & 0.382\\
\botline
\end{tabular*}
\end{table} 

The primary performance statistics at a 10-min lead time are presented in Table 2. The AUC score is the area under the receiver operating curve \citep[ROC;][]{Mason1982} and assesses a model’s ability to discriminate between classes. An AUC score of 0.5 indicates a no-skill forecast model, while a score of 1.0 is an indication of a perfect discriminator. The ResNet substantially outperforms the baseline logistic regression model in terms of the AUC score (Table 2). While the AUC score is considered overly optimistic and less informative for rare event predictions because it weighs positive and negative events equally \citep{Flora2021,Leinonen2022}, it’s still a rather useful metric for our study because our testing dataset has been downsampled to be balanced. ROCs of POD versus POFD, which are a measure of forecast skill at different probability thresholds at lead times of 10, 30, and 60 minute, are illustrated in Fig. 3a. Across all probability thresholds and lead times, the Resnet consistently performs better than the logistic regression model, with the magnitude of their differences diminishing as lead time increases. At the 10-min lead time, the maximum value of the ResNet Pierce skill score (PSS; defined as POD-POFD) is indicated by the red star in Fig. 3a, where the POD is higher than 0.85 and the POFD is lower than for the logistic regression model. 

\begin{figure*}[t]
\centerline{\includegraphics[width=\textwidth,angle=0]{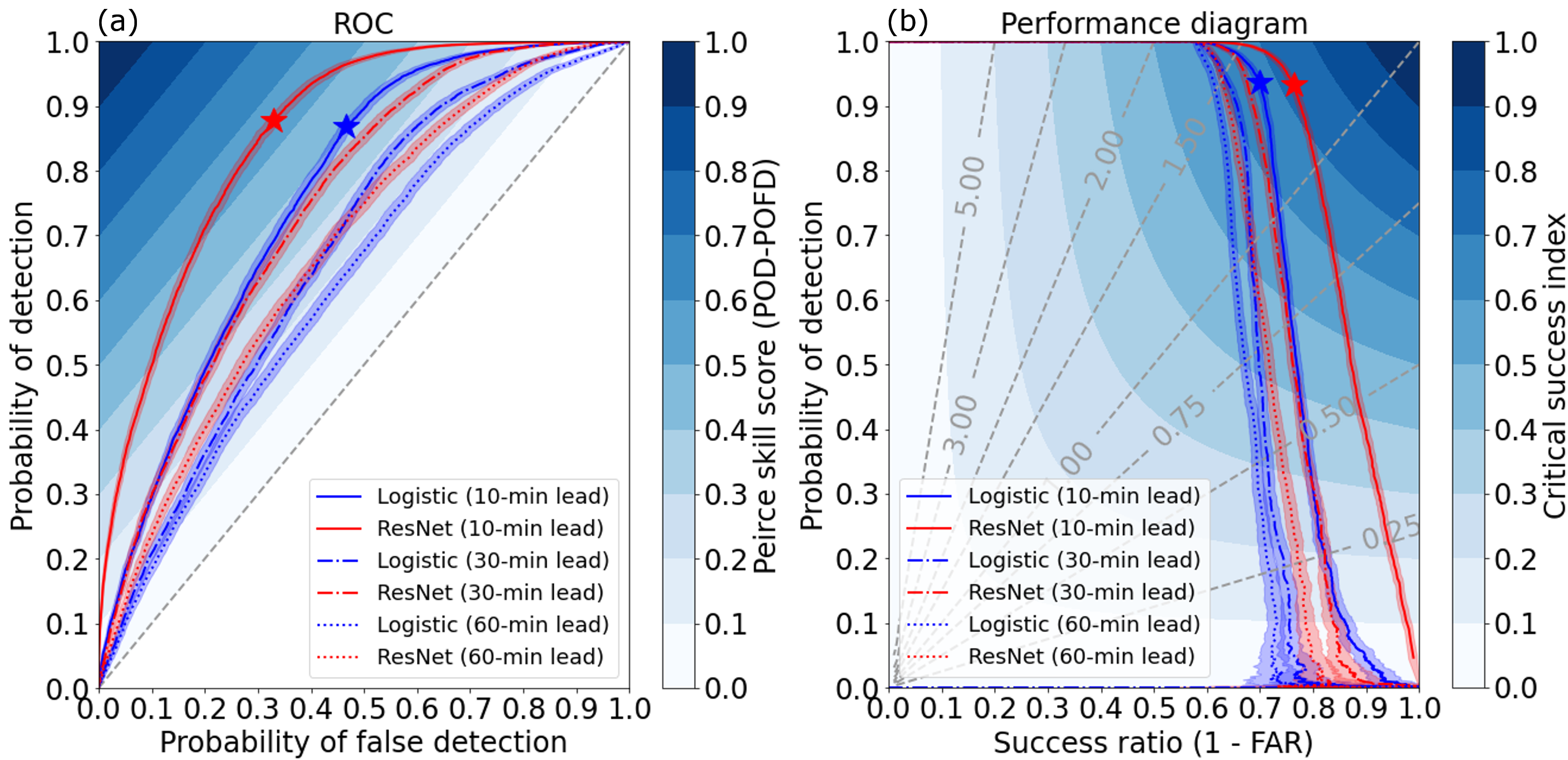}}
\caption{Performance of the logistic regression (blue) and ResNet (red) models on the testing dataset at lead times of 10 (solid), 30 (dashdotted), and 60 (dotted) minutes. Each curve shows the means for thresholds ranging from 0 to 1. Light shading around each line shows the 95\% confidence intervals determined by bootstrapping the testing samples 1000 times. (a) ROCs with the diagonal dashed line indicating a no-skill random classifier. Filled contours are the Pierce skill scores (PSS; defined as POD - POFD). For the 10-min lead time, the maximum PSS is marked by a star on each curve. (b) Performance diagrams with the dashed lines representing the frequency bias. Filled contours are the critical success indices (CSIs). For the 10-min lead time, the threshold that maximizes the CSI on the validation data is marked by a star on each curve.}\label{fig3}
\end{figure*}

\begin{figure*}[t]
\centerline{\includegraphics[width=\textwidth,angle=0]{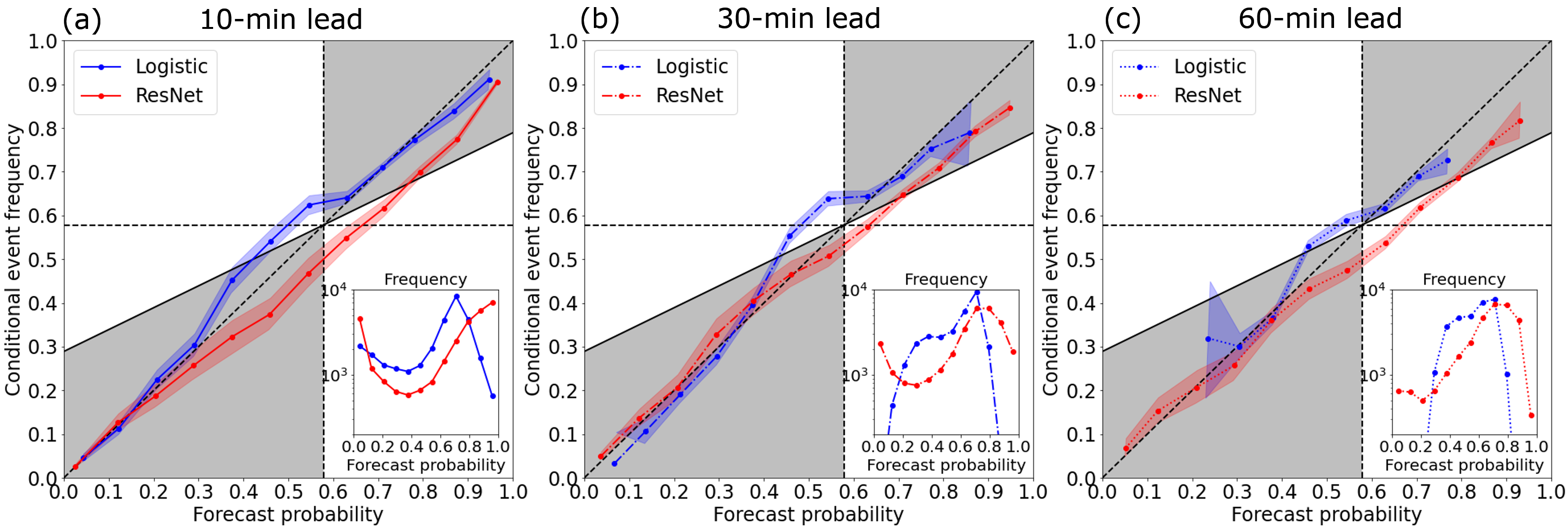}}
\caption{Attribute diagrams for the logistic regression (blue) and ResNet (red) models on the testing dataset at lead times of 10 (a), 30 (b), and 60 (c) minutes. Each curve shows the means achieved over all thresholds from 0 to 1. Light shading around each curve shows the 95\% confidence intervals determined by bootstrapping the testing samples 1000 times. The diagonal dashed line indicates perfect reliability, and the horizontal dashed line represents the climatological event frequency. The gray shaded areas indicate regions where points on the curves produce positive BSSs, whereas the white areas indicate regions where points on the curve generate negative BSSs. The inset panel shows the binned frequencies of the forecast probabilities for each model.}\label{fig4}
\end{figure*}

Forecast skill is further evaluated with a performance diagram (Fig. 3b) at lead times of 10, 30, and 60 minute. The performance diagram contains the POD versus the SR, thereby emphasizing a model’s ability to predict positive events while ignoring correct negative events \citep{Roebber2009}. Frequency bias (gray dashed lines) and CSI (filled contours) are also displayed in the performance diagram. Frequency bias, the ratio of total positive forecasts to total positive events, is a measure of bias resulting from class imbalance; a balanced dataset has a value of 1. The CSI is a significant metric for severe weather prediction because events like CI and tornado occurrences hold greater importance than non-events. For the 10-min lead time, the optimal threshold that maximizes CSI (Fig. 3b, stars) on the validation dataset almost maximizes CSI on the testing dataset for both models, suggesting that general characteristics, like class proportion and input feature distribution, of the validation and testing datasets are highly consistent. At the optimal threshold, the ResNet demonstrates a POD above 0.90, which is significantly higher than the SR. This indicates that the ResNet model's performance is primarily influenced by false alarms rather than misses. \cite{Brooks2018} have found that for rare event forecasts, achieving a certain amount of decrease in false alarms requires a much larger increase in misses, consistent with our results. Given that both ROC and performance diagram show the separation between the logistic regression and ResNet models, we can conclude that the ResNet is utilizing additional information in the data that the logistic regression couldn't use. The ResNet produces a higher CSI with a lower FAR than the baseline logistic regression model (Table 2) at the 10-min lead time. As the lead time increases, the advantage of ResNet over logistic regression gradually decreases. High FAR is a noticeable issue in previous satellite-based CI nowcasting algorithms \citep{Mecikalski2015}. These results indicate that the additional complexity of the ResNet model encodes localized spatial features that help to reduce FAR.   

\begin{figure*}[t]
\centerline{\includegraphics[width=25pc,angle=0]{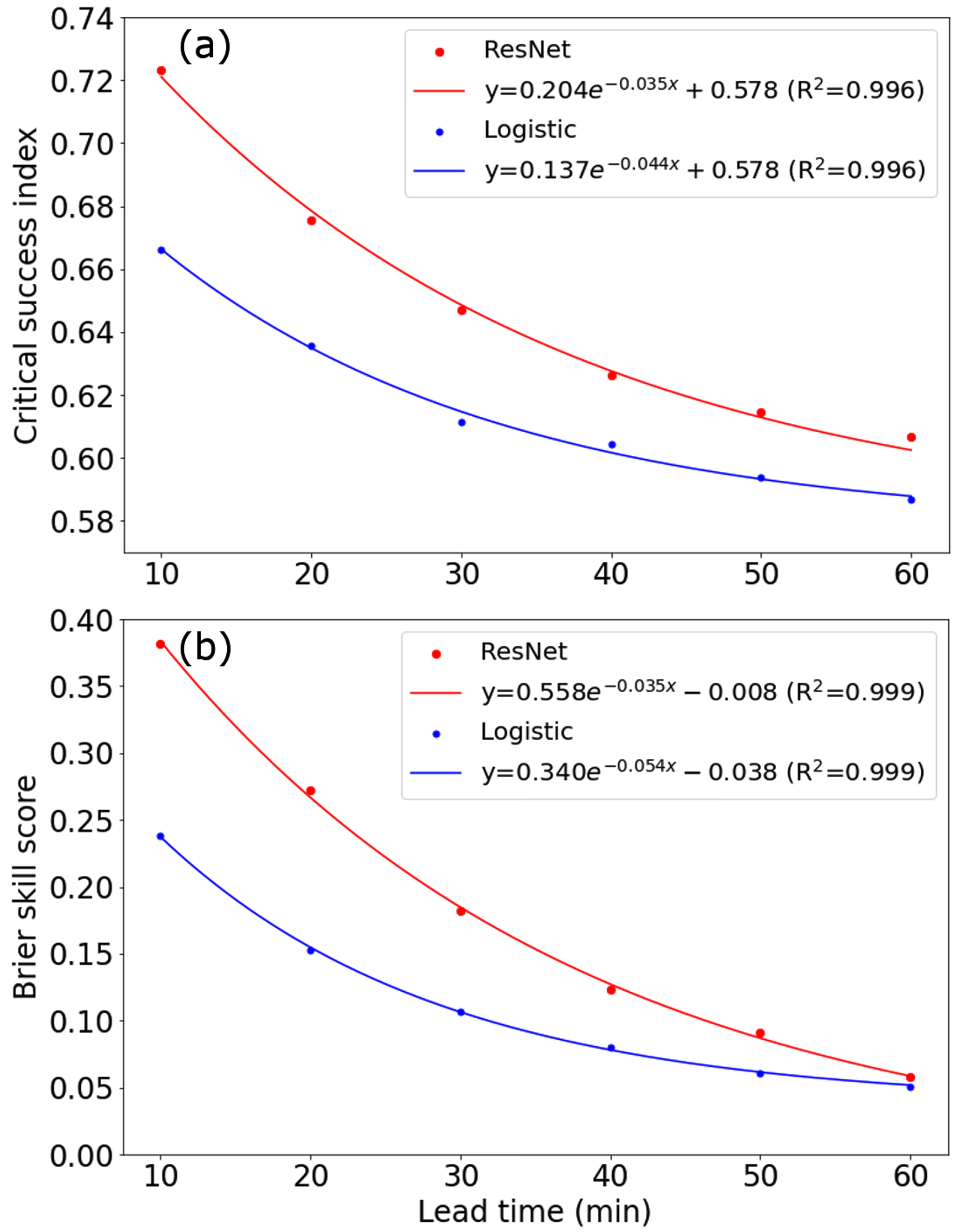}}
\caption{(a) CSIs and (b) BSSs for the ResNet (red) and logistic regression (blue) models at lead times from 60 min to 10 min in 10-min steps. Each solid line represents an exponential fit using $y=ae^{-\lambda x}+b$ ($b$ is preset to 0.578 for fitting CSI). R-squared ($R^2$) values are also provided to assess the goodness of fit.}\label{fig5}
\end{figure*}

The BSS of ResNet is better than for the logistic regression model (Table 2) at the 10-min lead time. Elements of the reliability and resolution terms in the BSS at lead times of 10, 30, and 60 minutes are illustrated in the attribute diagrams of Fig. 4, which shows the conditional event frequency against the forecast probability. At all lead times, the reliability curves of both models are close to the perfect reliability curve (Fig. 4, diagonal dashed line), and thus their differences in the reliability term are relatively small (< 0.004). Consistently across all lead times, the ResNet has an over-forecasting bias at probabilities over 0.4, whereas the logistic regression model consistently displays an under-forecasting bias at probabilities around 0.4. At the 10-min lead (Fig. 4a), the ResNet has more forecasts at probabilities less than 0.2 and greater than 0.8 compared to the logistic regression model, leading to a higher resolution with a difference around 0.04 compared to the logistic regression model. Thus, the higher BSS of ResNet at the 10-min lead time is mainly attributable to the sharper forecast probability distribution than for the logistic regression model. As the lead time increases (Fig. 4b-c), the number of forecasts from both models at probabilities less than 0.2 and greater than 0.8 decreases, signifying a reduction in resolution at longer lead times.

The skills of both models are further evaluated at additional lead times. Figure 5 shows CSIs and BSSs at lead times from 60 min to 10 min in 10-min steps along with exponential fits $y=ae^{-\lambda x}+b$ to them. The performances of both models degrade with increasing lead time as expected. For both models, the high correlation coefficients between both CSIs and BSSs and their exponential fits indicate that both scores decrease exponentially with lead time. With increasing lead time, CSIs for both models tend asymptotically to a lower limit close to the CSI (0.578) of climatological forecast using the event frequency of 0.578. BSSs for both models approach 0 asymptotically with increasing lead time, suggesting that both models degrade towards climatology. ResNet decreases much faster than logistic regression for both scores, and their differences largely disappear at longer lead times. Thus, local spatial features, as the major advantage of the ResNet, likely become less important with increasing lead time.

\subsection{Example cases}
We chose three characteristic examples of good and poor forecasts from the ResNet at lead times from 60 min to 10 min. The ResNet model, as the examples will show, is sensitive to a variety of spatial features, including water vapor amounts and the location of clouds. The probabilistic predictions of the logistic regression model are also included to explain the skill differences between the ResNet and the logistic regression models at different lead times.

\begin{figure*}[t]
\centerline{\includegraphics[width=\textwidth,angle=0]{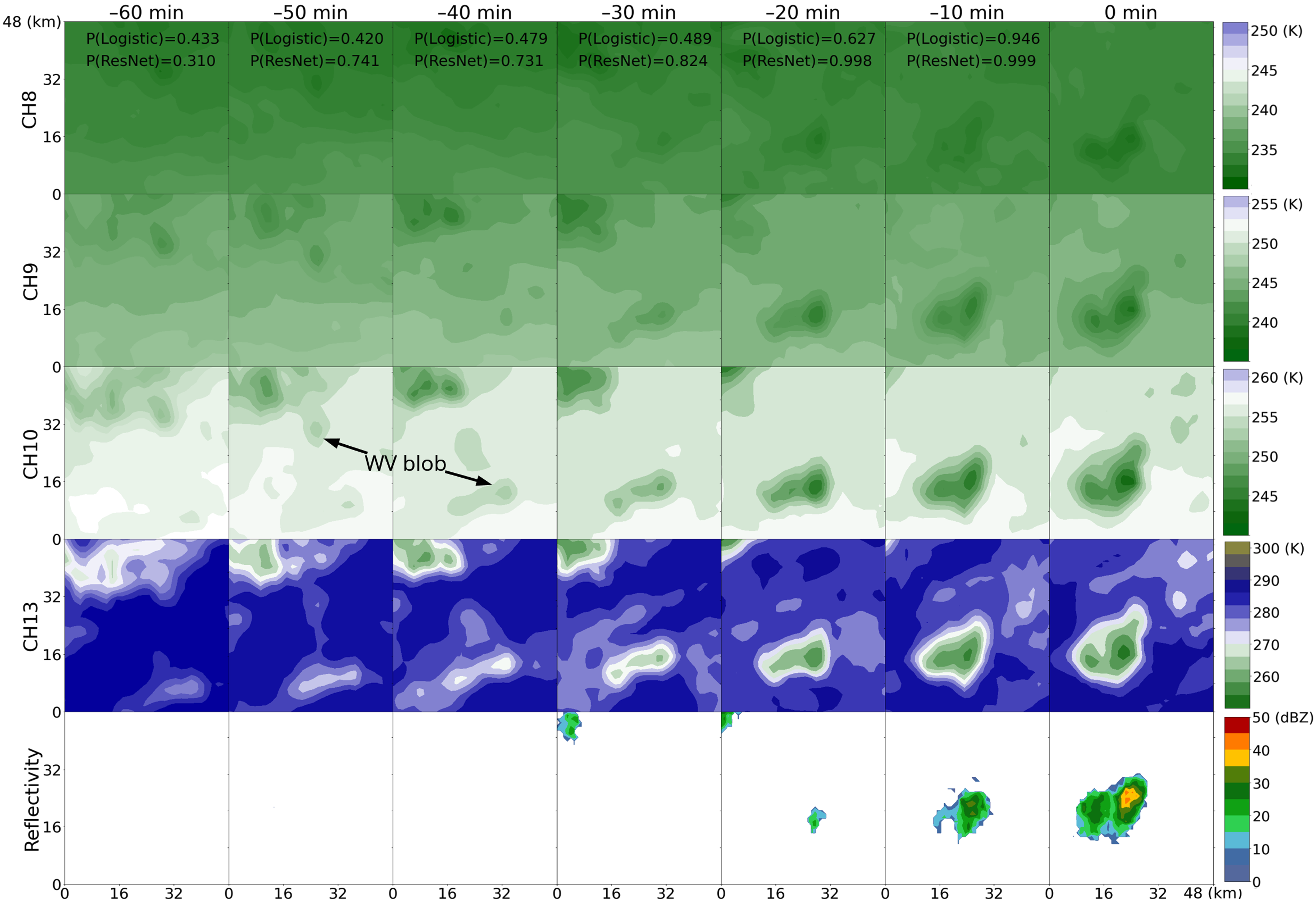}}
\caption{An example of a growing cloud object matched to a CI event at 0756 UTC 6 June 2021 with high probabilities for the ResNet and the logistic regression models at a 10-min lead time. The top four rows of images show the observed GOES-16 channel 8, 9, 10, and 13 brightness temperatures, respectively, at lead times indicated on the top of each column. The bottom row of images shows the observed MRMS composite reflectivity. The probabilistic predictions of the ResNet and the logistic regression models at different lead times are indicated just underneath the lead times at the top. The annotation highlights the two water vapor (WV) blobs associated with the cloud object.}\label{fig6}
\end{figure*}

\begin{figure*}[t]
\centerline{\includegraphics[width=\textwidth,angle=0]{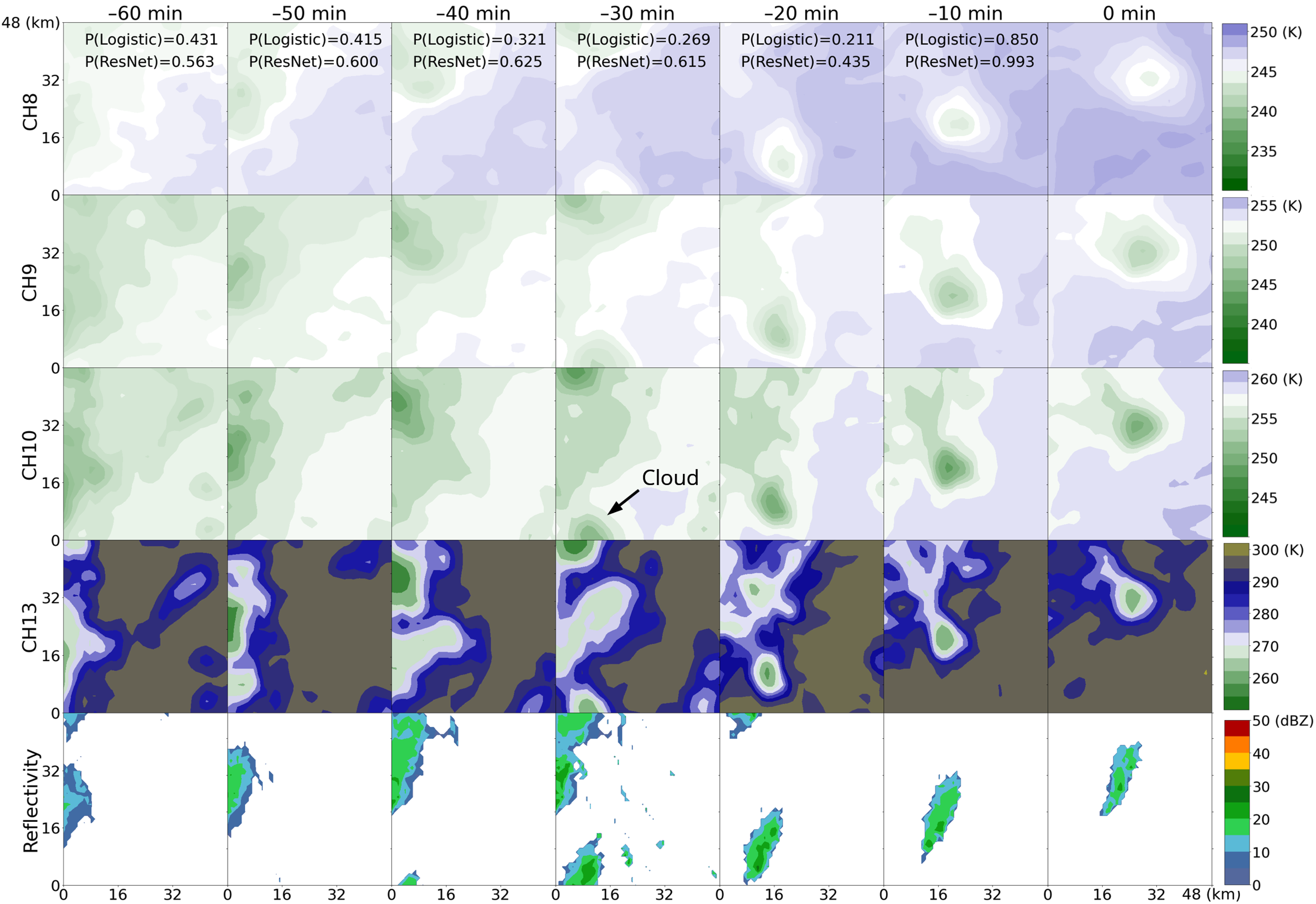}}
\caption{As in Fig. 6, but for a moving cloud object not matched to any CI event at 1942 UTC 25 June 2021 with high probabilities for both models at a 10-min lead time. The annotation highlights the preexisting cloud object associated with high probabilities for CI.}\label{fig7}
\end{figure*}

\begin{figure*}[t]
\centerline{\includegraphics[width=\textwidth,angle=0]{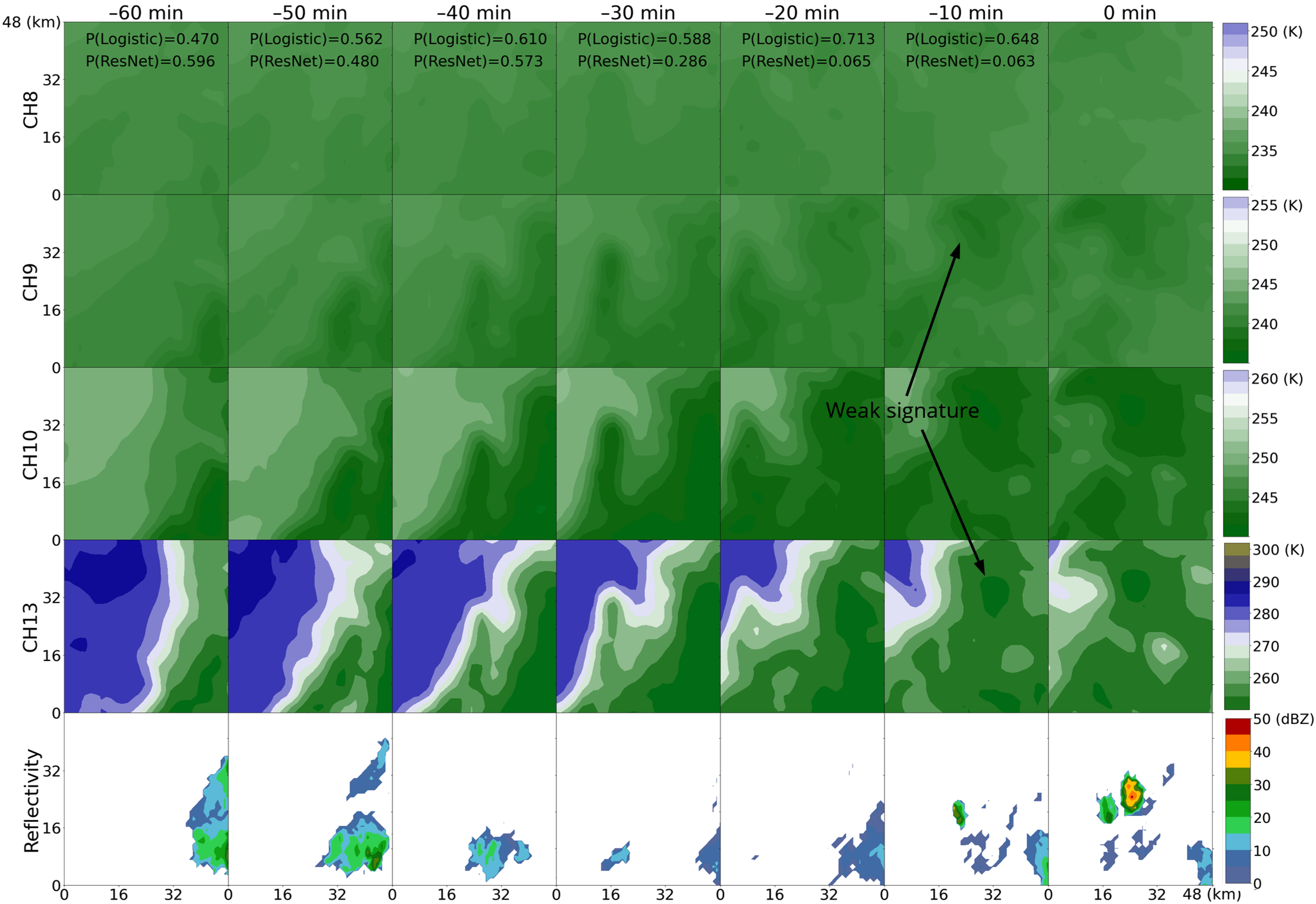}}
\caption{As in Fig. 6, but for a CI event at 0228 UTC 28 June 2021 obscured by cirrus clouds with a low probability for the ResNet and a high probability for the logistic regression model at a 10-min lead time. The annotation highlights the weak signature associated with the upcoming CI beneath the cirrus cloud.}\label{fig8}
\end{figure*}

The first example in Fig. 6 shows a growing cloud object matched to a CI event with high probabilities (close to 1) for both models at lead time t = $-$10 min. The GOES-16 channel 8, 9, and 10 brightness temperatures contain information about water vapor at upper ($\sim$344 hPa), middle ($\sim$442 hPa), and lower ($\sim$618 hPa) tropospheric levels, whereas the channel 13 brightness temperatures contain information on cloud-top temperatures, hence cloud-top heights. CI probabilities for both models are low at t = $-$60 min likely because of the relatively homogeneous water vapor amounts in the central region. The CI probability for the ResNet increases over 0.7 when the initial water vapor blobs of the cloud object first appeared in the lower troposphere at t = $-$50 min and $-$40 min. From t = $-$40 min to $-$20 min, the probability for the ResNet increases gradually to 0.998, while moisture convergence in the lower troposphere under the cloud object (i.e., the object bounded by white in the channel 13 images) likely intensified and transported hydrometeors upwards to colder temperatures. At t = $-$20 min, the much cooler cloud-top brightness temperatures of the cloud object indicate the occurrence of hydrometeors here, consistent with the observed 20-dBZ reflectivity. From t = $-$20 min to t = $-$10 min, the cloud expanded, moved towards the center, and became deeper with more moisture transported from the lower troposphere to the upper troposphere. The location and coverage of the clouds shown in the infrared observations match well with the radar reflectivity observations. The example demonstrates that the ResNet is sensitive to the location, height, and coverage of the clouds, and water amounts at different heights, especially in the upper troposphere. Before condensed water forms, the ResNet is perhaps using features indicating moisture convergence to predict CI at t = $-$50 min. In contrast, CI probability for the logistic regression model is less than 0.5 for lead times from 60 min to 30 min, quickly increasing from t = $-$30 min to t = $-$10 min. Thus, the logistic regression model is likely sensitive to the lowest brightness temperatures in all channels and the number of cold brightness temperatures within the cloud object.

In contrast to the first example, Fig. 7 shows a false alarm example of a moving cloud object not matched to any CI event with high probabilities for both models at a 10-min lead time. Between t = $-$60 min and t = $-$30 min, a preexisting cloud covered the upper left quadrant of the scene, corresponding to moderate CI probabilities around 0.6 for the ResNet. During the period, the environment in the middle and upper troposphere is much drier than for the first example, consistent with the lower probabilities for both models compared to the first example. From t = $-$30 min to t = $-$10 min, a cloud object moved from the bottom left of the scene to its center. The high probability for the ResNet at t = $-$10 min indicates the ResNet is sensitive to the location of clouds. The low probability for the ResNet at t = $-$20 min might be associated with the limited cloud coverage and the location of the cloud away from the center. The increase of the probability for the logistic regression model from t = $-$20 min to t = $-$10 min indicates that it might be sensitive to the lowest brightness temperatures in the lower troposphere. The slight decrease in the cloud-top temperature of the preexisting cloud from t = $-$20 min to t = $-$10 min in channel 13 implies that temporal variations, not encoded in this study, might be useful for reducing false alarms for this and similar cases.

Unlike the previous two examples, Fig. 8 shows a miss example of a CI event obscured by cirrus clouds. The cirrus anvil from a preexisting storm obscured a large region of the scene starting from the bottom right at t = $-$60 min. Only weak signatures of moisture convergence localized water vapor accumulation and developing clouds are evident in the middle troposphere and at cloud top at t = $-$10 min. These signatures are much weaker compared to the previous two examples. The CI probability for the ResNet is below 0.3 from t = $-$30 min to $-$10 min. Interestingly, the logistic regression model produced high CI probabilities, over 0.6, from t = $-$30 min to $-$10 min. Consistent with our interpretation of the first example, the logistic regression model is likely sensitive to the lowest brightness temperatures and the number of cold brightness temperatures. In pre-CI environments, it’s common for growing cumulus clouds in the lower troposphere to be obscured by cirrus anvils from pre-existing convection (Mecikalski et al. 2013). However, previous CI nowcasting algorithms (Mecikalski et al. 2015; Apke et al. 2015) focused on predicting CI for cumulus cloud objects identified from satellite observations, ignoring CI events whose radiative signatures were partially or completely obscured by cirrus clouds. While the ResNet is less skillful for the cases obscured by cirrus clouds, this example indicates that temporal variations of cloud-top brightness temperature might be essential for enhancing the predictive skill of similar cases.

These manually selected examples do not cover the wide variety of cases in the testing dataset and the real world. However, they demonstrate how the models respond to different environments at different lead times and how to overcome clear limitations in the current methodology, thereby motivating specific improvements in future work.

\section{Model explanation}

\begin{figure*}[t]
\centerline{\includegraphics[width=\textwidth,angle=0]{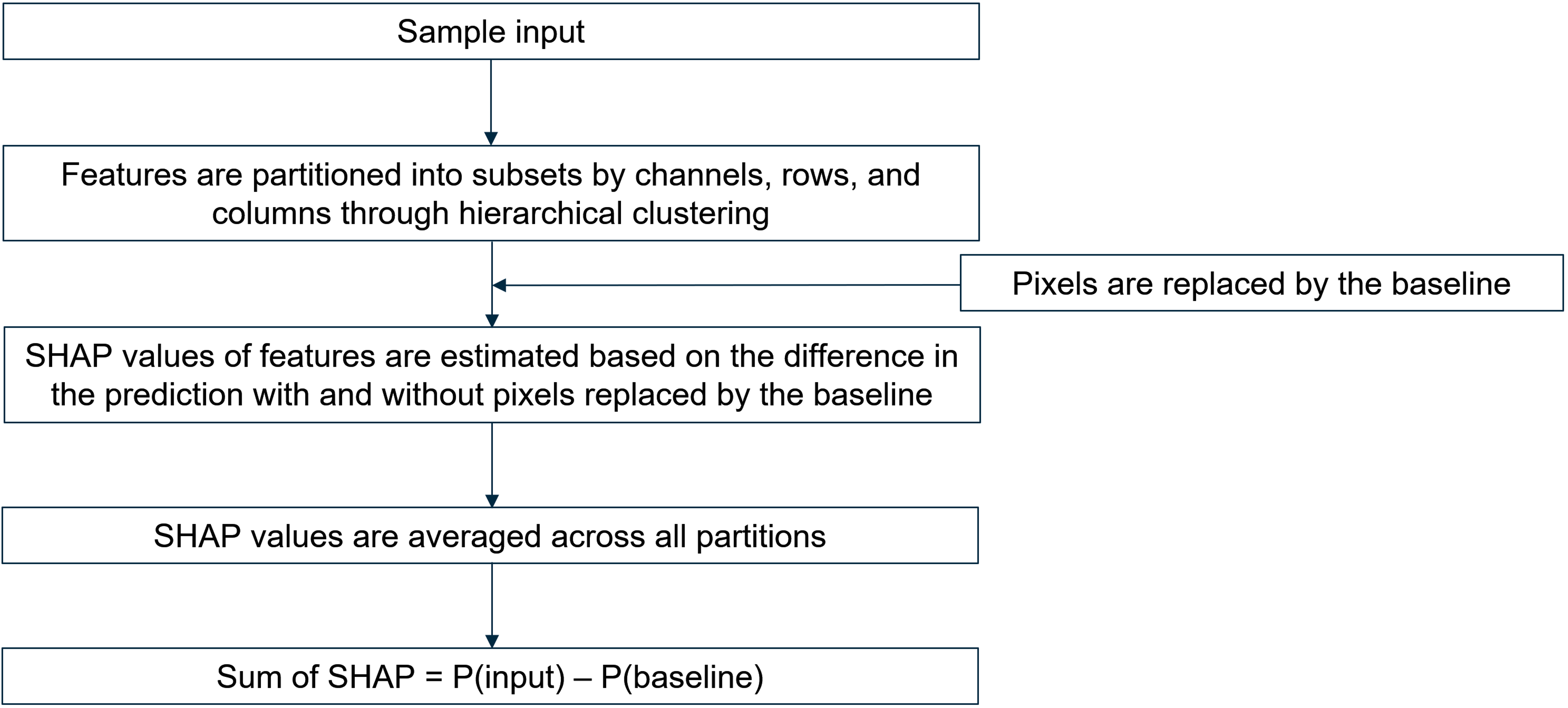}}
\caption{Flowchart for the PartitionSHAP method. Each box represents an individual action.}\label{fig9}
\end{figure*}

In order to explain features encoded within the ResNet model, SHAP values were estimated using PartitionSHAP by \cite{Krell2021}. The procedure of PartitionSHAP is shown in Figure 9. PartitionSHAP applies feature partitioning to explain the contribution of features to the prediction. For each sample input, the feature space is first partitioned into subsets by channels, rows, and columns using hierarchical clustering that groups similar features into clusters. For each feature subset, the SHAP values, a measure of the contribution to the prediction, are estimated for it while considering interactions within the subsets via the difference between the model’s predictions for a specific sample with and without pixels replaced by the baseline. Then, the SHAP values are averaged across partitions to obtain a single set of SHAP values as the measure of the feature attribution. Given the additivity property of SHAP values, the sum of SHAP values for all features approximates the prediction difference between the sample input and the baseline. Features with higher positive SHAP values have a larger positive contribution to the prediction difference, while features with higher negative SHAP values have a larger negative contribution to the prediction difference.

\begin{figure*}[t]
\centerline{\includegraphics[width=\textwidth,angle=0]{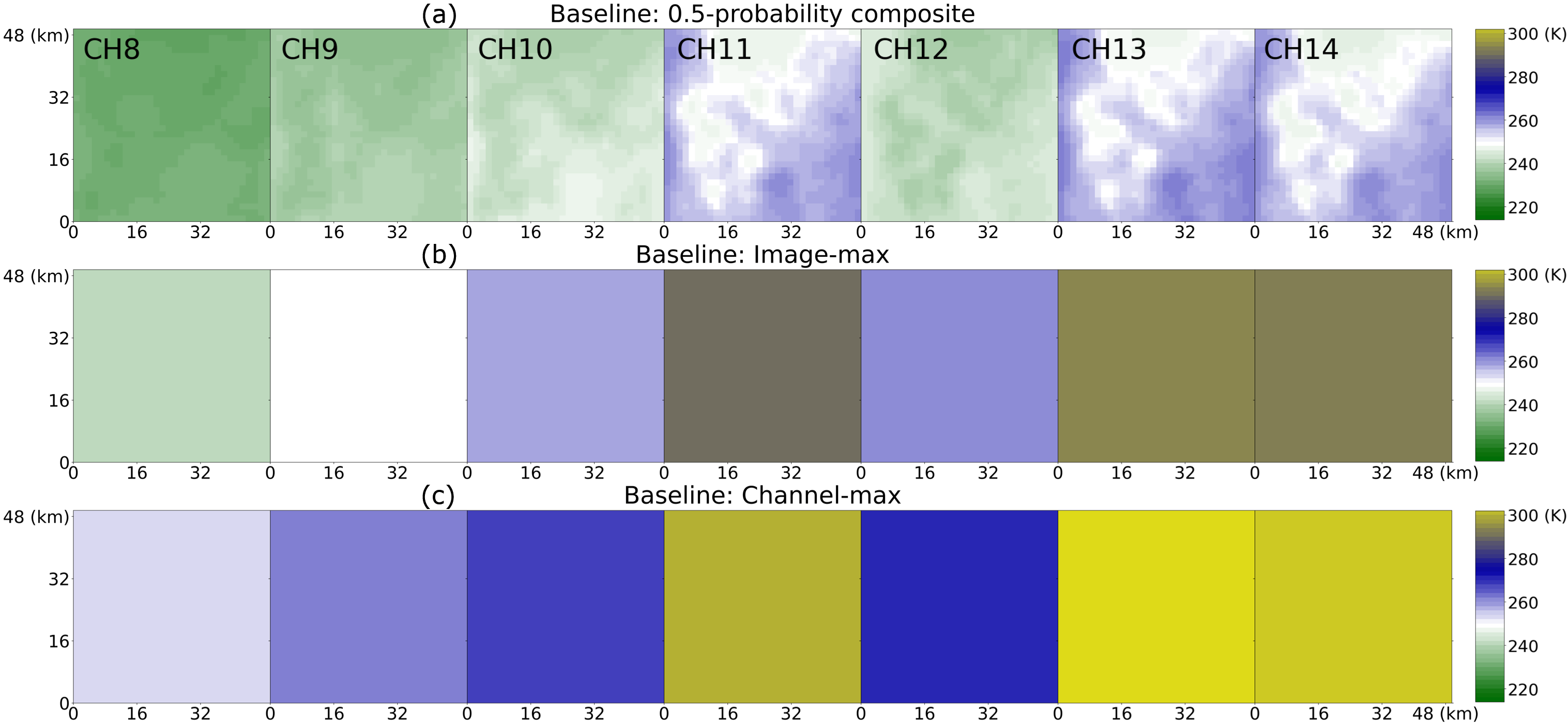}}
\caption{Baselines used in SHAP calculations. (a): 0.5-probability composite baseline is the PMM composite of 100 0.5-probability samples with a specific percentile range from 36.6\% to 63.4\% to uphold the composite’s 0.5 probability. (b): The ensemble mean of the 100 image-max baselines for the 100 cases. For each case, the image-max baseline comprises seven 32×32 channel-specific uniform arrays representing the maximum brightness temperature found in the image of each channel. (c): Channel-max baseline comprises seven 32×32 channel-specific uniform arrays representing the maximum brightness temperature found across all 100 images for each channel.}\label{fig10}
\end{figure*}

\begin{figure*}[t]
\centerline{\includegraphics[width=\textwidth,angle=0]{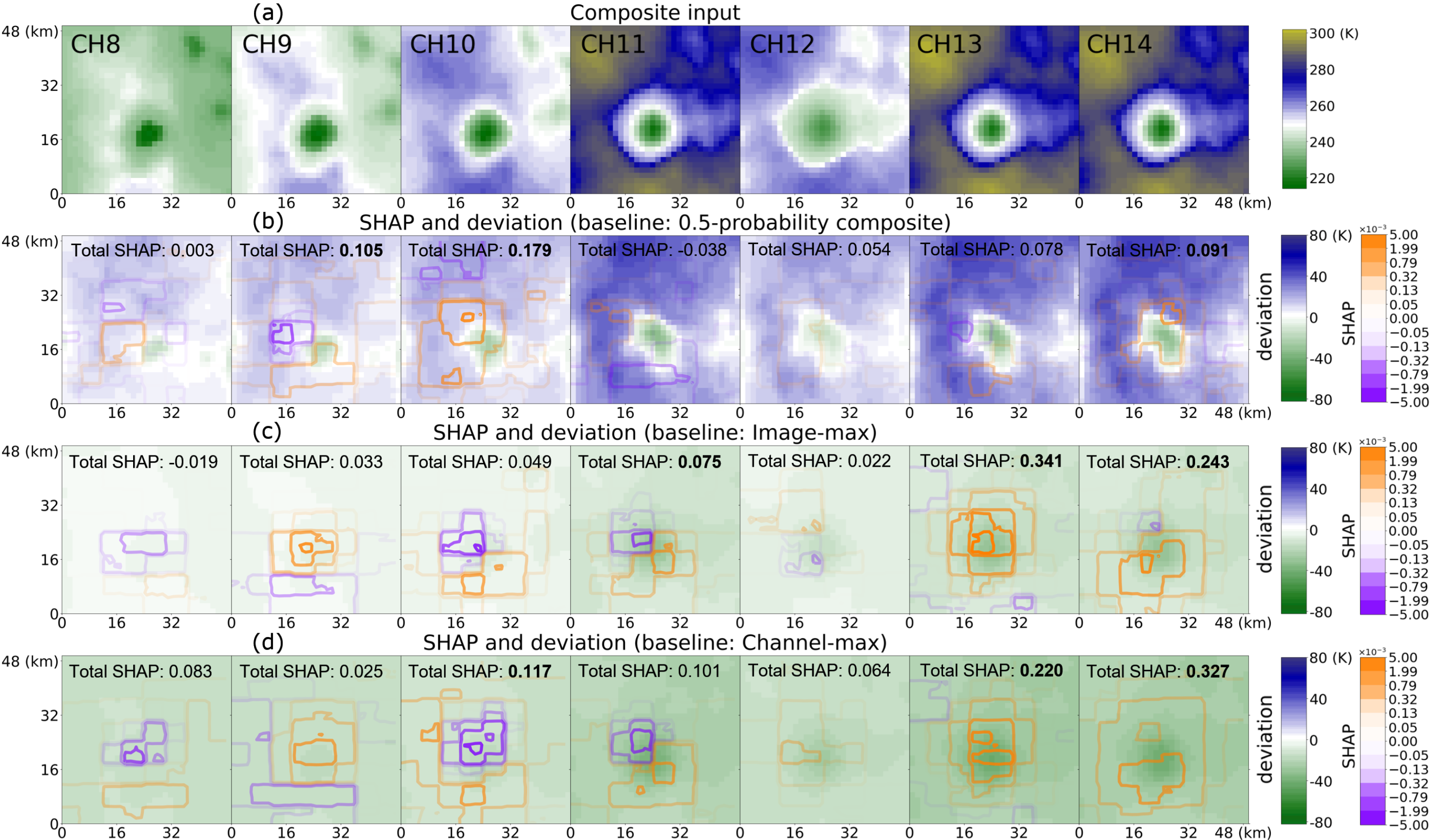}}
\caption{(a): PMM brightness temperature composite inputs for channels 8 through 14 generated from the 100 best hit cases at a 10-min lead time. (b-d): Composite SHAP heat maps (contours) and composite deviations (shading) of the inputs from the (b) 0.5-probability composite, (c) image-max, and (d) channel-max baselines. All of the composites are created by applying the PMM method to the images from the 100 best hit cases. Positive SHAP values indicate positive contributions to the prediction difference between the input and the baseline, whereas negative SHAP values indicate negative contributions. The total SHAP of each channel is indicated at the top of each image. The top three maximum total SHAPs for each baseline across channels are highlighted in bold.}\label{fig11}
\end{figure*}

SHAP output is a heat map overlaid on the deviation of the input from the baseline to reveal the additive contribution of each pixel to the prediction difference between the input and the baseline. The baseline is a reference against which changes in predictions using sample inputs are interpreted by comparison. \cite{Mamalakis2022} demonstrated that model explanation is highly dependent on the baseline and different baselines can be used to answer different science questions. We explored the important radiative features behind the 100 best hit cases with CI probabilities close to 1 using three different baselines (Fig. 10). The first baseline is a composite of 0.5-probability samples (Fig. 10a, hereafter 0.5-probability composite). Here, we used a probability-matched mean \citep[PMM;][]{Ebert2001} composite of 100 0.5-probability samples with a specific percentile range of brightness temperatures from 36.6\% to 63.4\% to uphold the composite’s 0.5 probability. PMMs preserve spatial structures better than simply taking the mean of inputs. The slightly depressed brightness temperatures near the center of Fig. 10a are consistent with a moderate CI probability generated by the ResNet. We call the second baseline image-max baseline (Fig. 10b). For each case, the image-max baseline comprises seven 32×32 channel-specific uniform arrays representing the warmest brightness temperature found in the image of each channel. Figure 10b shows the ensemble mean of the image-max baselines over the 100 cases. Finally, the channel-max baseline (Fig. 10c) comprises seven 32×32 channel-specific uniform arrays representing the warmest brightness temperature found across all 100 images for each channel.

Figure 11 shows composite inputs of the 100 best hit cases (Fig. 11a) and the SHAP and deviation values for the three baselines (Fig. 11b-d). For each baseline, the top three maximum total SHAPs across the seven channels are highlighted in bold, indicating the channels that receive the highest attention from the ResNet model. The best hits composite (Fig. 11a) is characterized by a strong blob near the center with moisture accumulated in the lower (CH10) and middle (CH9) troposphere, localized moisture accumulation in the upper troposphere (CH8), and cloud coverage observed in the window channels (CH13/14). With the 0.5-probability composite baseline, the question to be addressed is as follows: “Which features made the model predict CI compared to a relatively moist environment with a weak blob near the center and a moderate CI probability?” Based on SHAP results (Fig. 11b), CI probability mainly comes from moisture gradients between the blob and the environment in the lower troposphere (CH10), moisture gradients surrounding the blob in the middle troposphere (CH9), and the brightness temperature gradient at cloud boundaries observed in CH14. Negative contributions are mainly from the drier areas in the middle troposphere near the center. Thus, the results indicate that the model has learned that relative to the 0.5-probability composite baseline, CI is mostly determined by moisture gradients near the blob, possibly associated with moisture convergence, in the lower and middle troposphere.

We then use the image-max baseline to answer the following question: “Which features made the model predict CI compared to a relatively dry environment?” The SHAP results (Fig. 11c) highlight positive contributions mainly from the cloud-top height observed in CH13, brightness temperature gradient at cloud boundaries in CH14, and cloud-top phase changes near the blob in CH11. While weak positive contributions arise from moisture gradients in the vicinity of the blob in the lower troposphere (CH10), they are largely counterbalanced by negative contributions from the central regions. These negative SHAP might arise from the model's knowledge gained from other CI cases, especially those obscured by cirrus clouds. With the image-max baseline, the ResNet is more focused on cloud-top height and phase changes as well as cloud coverage.

Model explanation is further explored with the channel-max baseline to gain insights on the following question: “Which features made the model predict CI relative to a dry environment?” The SHAP results (Fig. 11d) highlight the positive contributions from cloud-top height (CH13/14), cloud coverage (CH13/14), and the moisture environment in the lower troposphere (CH10). More positive contributions stem from the moisture within the lower troposphere compared to the results with the image-max baseline, possibly attributed to a generally higher contrast of moisture content in the environment. The comparison demonstrates that the model prediction is not only dependent on the characteristics of the blob and cloud objects, but also takes environmental moisture into account. 


\begin{figure*}[t]
\centerline{\includegraphics[width=\textwidth,angle=0]{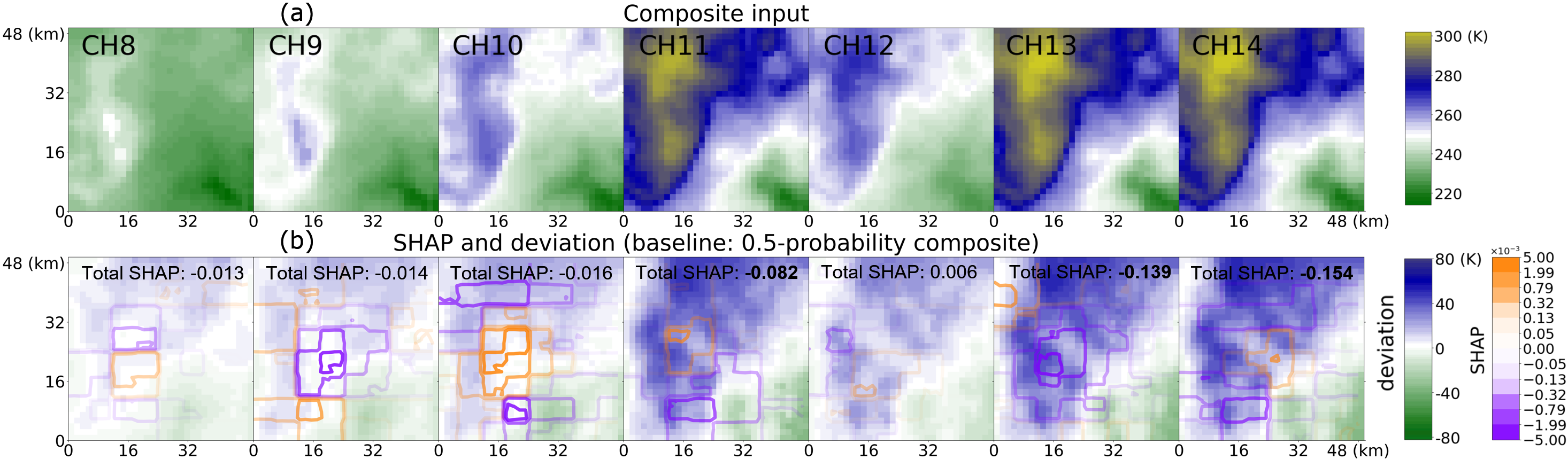}}
\caption{(a) PMM composite of inputs over 100 worst miss cases and (b) the SHAP and deviation values relative to the 0.5-probability composite baseline.}\label{fig12}
\end{figure*}

We further explored the model explanation using the 0.5-probability composite baseline on the 100 worst miss cases. We aim to understand why the model missed some CI events through the SHAP analysis. Figure 12 displays the PMM composite of inputs over the 100 worst miss cases and the SHAP and deviation values from the 0.5-probability composite baseline. According to the composite input (Fig. 12a), areas in the upper-left are in clear-sky conditions whereas in the lower-right area it is cloudy. These conditions might obscure signals like moisture gradients in the lower and middle troposphere. According to the SHAP and deviation values (Fig. 12b), the model generated positive contributions from moisture gradients in the lower troposphere in the central region (CH10). Negative contributions stem from the wide clear-sky areas in the window channels (CH13/14) and the cloud-top temperature gradient near the cloud boundaries observed in CH11. Combined with the third case example (Fig. 8), we hypothesize that the ResNet fails to generate correct forecasts for these misses likely because the signatures in the middle and lower troposphere were blocked by the anvil clouds from preexisting convection. Surface observation data might help infer these missing signatures from convergence and differential heating near the surface \citep{Weckwerth2011}.

In the Appendix, we further explored the model explanation on worst false alarm and best correct null cases. The worst false alarms results (Fig. S1) are similar to the best hits results (Fig. 11), suggesting that the model might be unable to distinguish between advective clouds from the surroundings and convective clouds developed from lower-tropospheric blobs. The best correct null results are similar to the worst misses results (Fig. 12), still highlighting the negative contributions from the clear-sky regions.


\section{Discussion and Conclusions}
Convective initiation nowcasting from satellite observations has proved challenging for existing algorithms, yet this work demonstrates improvements in forecast skill and explainability. We presented a data-driven method for CI nowcasting at lead times up to 1 hour using a ResNet architecture for encoding spatial features of GOES-16 satellite observations. The ResNet model was compared against the classical logistic regression model to evaluate the improvements to skill added by spatial encodings. The ResNet model significantly outperforms the logistic regression model in multiple evaluation metrics at lead times up to 1 hour, especially on the false alarm ratio. However, improvements in prediction skill via encoding of spatial features quickly decreases with increasing lead time, indicating that spatial features associated with CI might be statistically weaker at longer lead times. Interestingly, the performance of both models decreases exponentially towards the climatology with increasing lead time. Through case studies, we found that the logistic regression model is sensitive to the lowest brightness temperatures and the number of cold brightness temperatures, whereas the ResNet model is sensitive to the location, height, and coverage of clouds, and moisture amounts at different levels. We also found that the ResNet model is unable to correctly forecast non-CI events associated with moving clouds or CI events whose signatures are obscured by overlying cirrus anvil clouds.

We suggest that model explanation answers different science questions based on the choice of different baselines. We employed the PartitionSHAP method to better estimate the contribution from feature interactions. With the 0.5-probability composite baseline, a moist baseline with a moderate CI probability, CI is mostly determined by moisture gradients near cold brightness temperature blobs, possibly associated with moisture convergence in the lower and middle troposphere. With the image-max baseline, the model focused attention on cloud-top height and phase changes as well as cloud coverage. With the channel-max baseline, contributions from cloud-top height, cloud coverage, and moisture environment in the lower troposphere were emphasized. Our study demonstrates the advantage of using different baselines in further understanding model decision-making processes and gaining potential scientific insights. The explanation results on 100 worst miss cases indicate that the failure of the model in these instances are likely caused by a lack of signatures in the lower and middle troposphere due to obscuration by preexisting clouds.

Though this work is but the first step in producing an operational CI forecasting system based on ML methods, it is an important one in demonstrating extraction of the physical processes encoded in the model that impact CI forecasting. Subsequent work will focus on incorporating predictors across multiple timesteps and additional meteorological information into CI nowcasting with an emphasis on skill, understanding the encoded physical processes, and operational resilience.

While these results are promising, there are some limitations that must be considered. First, because we are focused on predicting CI events against their neighboring non-CI events, we are not able to cover all types of non-CI events, including some cumulus cloud objects used in previous studies \citep{Mecikalski2015,Apke2015}. A benefit of our framework is that it includes CI events obscured by cirrus clouds which are ignored in previous work. Second, because non-CI events have been downsampled to be comparable to the number of CI events to make a balanced dataset, the class proportion of our dataset is different from realistic class proportions in the real atmosphere. Thus, performance evaluation against climatology, like the BSS score, might not be reliable. Third, the model explanation is still affected by  interactions between correlated features. Although PartitionSHAP was initially designed to better estimate contributions from the interaction between features, the results, especially the negative SHAP values, are still affected by interactions of localized features. Feature correlation might have been encoded into the model during the training. The model might have learned how to utilize the correlated features to maximize its skills. For example, the difference in brightness temperatures between CH11 and CH13 is usually used to provide information about cloud-top glaciation. ResNet might have encoded this signature in inferring the timing of CI. 


%

%

\clearpage
\acknowledgments
The authors thank David J. Stensrud and Yunji Zhang for discussions about CI definition and identification. This research benefits from NCAR’s Computational and Information Systems Laboratory visitor program. The machine learning training, evaluation, and explanation are performed on Institute for Computational and Data Sciences supercomputer provided by Penn State and the Cheyenne supercomputer (Computational and Information Systems Laboratory 2017), provided by the National Center for Atmospheric Research (NCAR)’s Computational and Information Systems Laboratory. This research was partially funded by the Penn State College of Earth and Mineral Sciences, a Penn State Institutes of Energy and the Environment seed grant, and the National Science Foundation under Grant No. ICER-2019758. This material is based upon work NCAR, which is a major facility sponsored by the National Science Foundation (NSF) under Cooperative Agreement No. 1852977.

%
%
\datastatement
The machine learning and analysis software used in this paper can be accessed in the CIML library available at \url{https://github.com/dxf424/CIML}. Processed training and testing data are available online (\url{https://doi.org/10.26208/6Y59-0R80}).

%
\newpage

\appendix
\setcounter{figure}{0}
\makeatletter 
\renewcommand{\thefigure}{S\@arabic\c@figure}
\setcounter{equation}{0}
\renewcommand{\theequation}{S\@arabic\c@equation}
\setcounter{table}{0}
\renewcommand{\thetable}{S\@arabic\c@table}
\setcounter{section}{0}
\renewcommand{\thesection}{S\@Roman\c@section}



%
\subsection{Storm-Tracking configuration}
The w2segmotionll algorithm, a WDSS-II executable, and the modified best-track algorithm are used for storm identification and tracking from radar MRMS radar dataset. Table S1 shows the configuration options for these algorithms.
\begin{table}[h]
\caption{Configuration options used for storm tracking and identification for w2segmotionll and post-event track correction for best-track.}\label{tableS1}
\centering
\begin{tabular*}{0.85\hsize}{@{\extracolsep\fill}ccc}
\topline
 Parameter & Flag & Option \\
\midline
\ & \textit{\textbf{Storm tracking and identification for w2segmotionll}} & \\
\ trackedProductName & -T & MergedReflectivity QCComposite\\ 
\ “min max incr maxdepth” & -d & “35 57 5 -1” \\
\ prunerSizeParameters & -p & 40, 200, 300, 0:0, 0, 0\\
\ smoothing filters & -k & percent:50:1:0:1, percent:75:1:0:1 \\
\ clusterIDMatchingMethod & -m & MULTISTAGE: 2:10:0 \\
\ & \textit{\textbf{Post-event track correction for best-track}} & \\
\ Buffer distance & -bd & 16 (km)\\
\ Buffer time & -bt & 11 (min)\\
\botline
\end{tabular*}
\end{table} 

\subsection{Hyperparameter tuning}
Table S2 shows the selected hyperparameters, their search space, and optimal values for the baseline logistic regression model and ResNet.

\begin{table}[h]
\caption{Selected hyperparameters, their search space, and optimal values for the baseline logistic regression model and ResNet.}\label{tableS2}
\centering
\begin{tabular*}{0.9\hsize}{@{\extracolsep\fill}ccc}
\topline
 Hyperparameter & Search space & Optimal value \\
\midline
\ & \textit{\textbf{Logistic regression}} & \\
\ C	 & 0.0001-1.0 (log-uniform) & 0.212 \\ 
\ $\lambda$ (mixing parameter) & 0.0001-1.0 (log-uniform) & 0.104 \\
\ & \textit{\textbf{ResNet}} & \\
\ Leaky alpha & 0.0-0.4 & 0.138\\
\ Learning rate & 0.0000001-0.001 (log-uniform) & 0.0009\\
\ Initial number of filters & 10-100 & 64\\
\ Batch size & 256-2048 & 256\\
\ Dropout alpha & 0.0-0.4 & 0.074 \\
\botline
\end{tabular*}
\end{table} 

\subsection{SHAP results for false alarms and correct nulls}
\begin{figure*}[h]
\centerline{\includegraphics[width=\textwidth,angle=0]{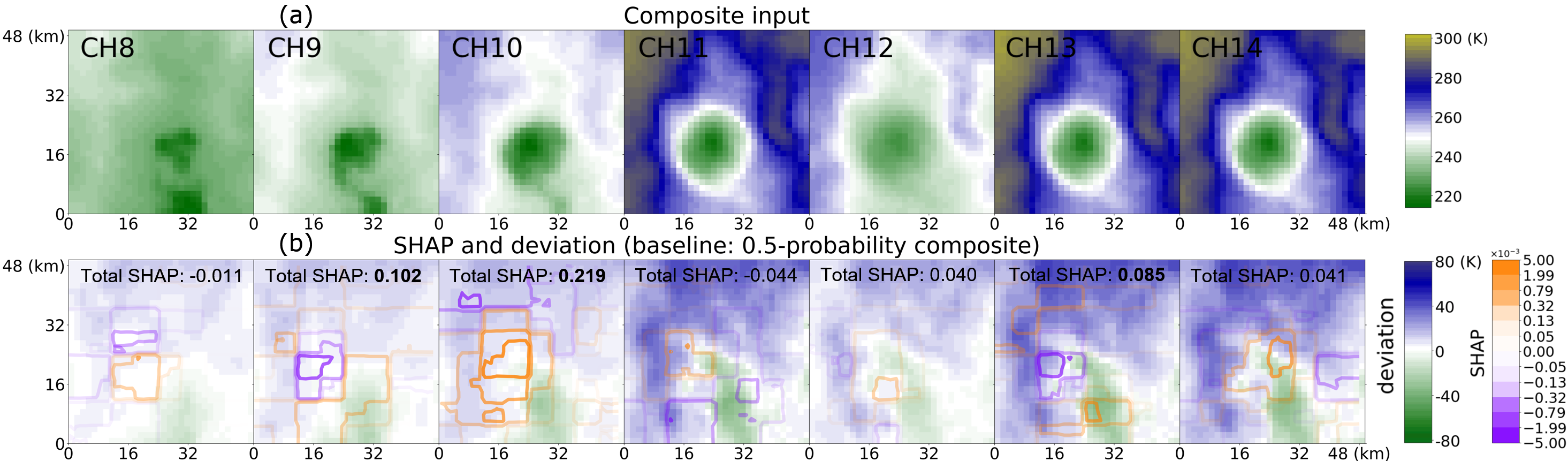}}
\caption{(a) PMM composite of inputs over 100 worst false alarm cases and (b) the SHAP and deviation values relative to the 0.5-probability composite baseline.}\label{figS1}
\end{figure*}

\begin{figure*}[h]
\centerline{\includegraphics[width=\textwidth,angle=0]{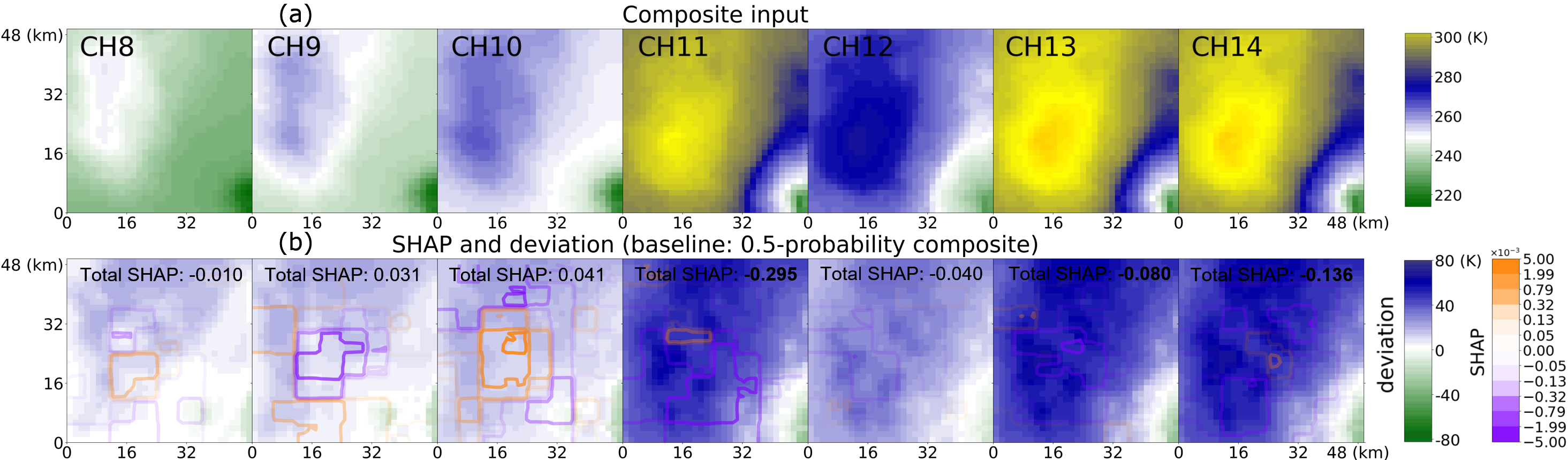}}
\caption{(a) PMM composite of inputs over 100 best correct null cases and (b) the SHAP and deviation values relative to the 0.5-probability composite baseline.}\label{figS2}
\end{figure*}

Figure S1 illustrates the composite inputs over 100 worst false alarm cases and SHAP results and deviation values relative to the 0.5-probability composite baseline. Similar to the best hits results (Fig. 11), the composite inputs of worst false alarms are characterized by moisture accumulated near the center in the troposphere as well as cloud coverage observed in the window channels (CH 13 and CH14). The SHAP results still highlights the positive contribution from the moisture gradients between the cloud object and the surrounding area in the lower troposphere (CH10) and middle troposphere (CH9) as well as the cloud-top height (CH13/14). The much wider cloud coverage than the cloud of best hit cases (Fig. 11a) indicates the cloud might be advected from the surroundings, consistent with the second case example (Fig. 7). Thus, the results suggest that the model might not be able to distinguish between advective clouds and convective clouds developed from lower-tropospheric blobs.

Figure S2 shows the composite inputs over 100 best correct null cases and SHAP results and deviation values relative to the 0.5-probability composite baseline. Similar to worst miss cases (Fig. 12), negative contributions are mostly from the wide clear-sky regions in CH11, 13, and 14.



\bibliographystyle{ametsocV6}
\bibliography{references}

\end{document}